\documentclass[aps,prd,preprint,showpacs,superscriptaddress]{revtex4}
\usepackage{graphicx}
\usepackage{bm}
\begin{document}

\title{Density field in extended Lagrangian perturbation theory}

\author{Takayuki Tatekawa}
\email{tatekawa@gravity.phys.waseda.ac.jp}
\affiliation{Department of Physics, Waseda University,
3-4-1 Okubo, Shinjuku-ku, Tokyo 169-8555, Japan}
%

\date{\today}

\begin{abstract}
We analyzed the performance of a perturbation theory for nonlinear
cosmological dynamics,
based on the Lagrangian description of hydrodynamics.
In our previous paper, we solved hydrodynamic equations for
a self-gravitating fluid with pressure, given by a polytropic
equation of state, using a perturbation method.
Then we obtained the first-order solutions in generic background
universes and the second-order solutions for a wider range of
polytrope exponents. Using these results, we describe density fields
with scale-free spectrum, SCDM, and LCDM models. Then we analyze
cross-correlation coefficient of the density field
between N-body simulation and Lagrangian
linear perturbation theory, and the probability distribution of density
fluctuation. From our analyses, for scale-free spectrum models,
the case of the polytrope
exponent $5/3$ shows better performance than the Zel'dovich approximation
and the truncated Zel'dovich approximation in a quasi-nonlinear regime.
On the other hand, for SCDM and LCDM models, improvement by the effect
of the velocity dispersion was small.
\end{abstract}

\pacs{04.25.Nx, 95.30.Lz, 98.65.Dx}

\maketitle

\section{Introduction}\label{sec:intro}

The Lagrangian approximation for structure formation in cosmological
fluids provides a relatively accurate model even in quasi-linear regime,
where density fluctuation becomes unity. The
Zel'dovich approximation (hereafter ZA)~\cite{zel,buchert92,coles,saco},
a linear Lagrangian approximation for dust fluid,
describes the evolution of density fluctuation better than Eulerian
approximation~\cite{munshi,sahsha}. Although ZA gives an accurate description
until quasi-linear regime develops,
ZA cannot describe the model after the formation
of caustics. In ZA, even after the formation of caustics, the fluid elements
keeps moving in the direction set up by the initial condition. ZA cannot
describe compact and high density structures such as pancakes,
skeletons, or clumps,
while N-body simulation shows the presence of clumps with a very wide range
in mass at any given time~\cite{davis}. In general, after the formation
of caustics, hydrodynamical description becomes invalid.

In order to proceed with a hydrodynamical description
without the formation of caustics, the qualitative pressure
gradient~\cite{zel82} and thermal velocity scatter~\cite{kotok,shandarin}
in a collisionless medium had been discussed.
After that, the `adhesion approximation'~\cite{gurbatov} was
proposed from the consideration of nonlinear wave equations like
Burgers' equation. In the `adhesion approximation', an artificial
viscosity term is added to ZA. As another modification,
a Gaussian cutoff is applied to the initial power spectrum of density
fluctuation and then evolves with the ZA. This modified approximation
is called the `Truncated Zel'dovich approximation' (hereafter TZA)
~\cite{cms,mps}.
These modified approximations successfully avoid the formation
of caustics and describe the evolution for long time.
However, the physical origin of the modification is not clarified.

We reconsider the basic, fundamental equation for the motion of matter.
The collisionless Boltzmann equation~\cite{BT} describes the motion
of matter in phase space. The basic equations of hydrodynamics
are obtained by integrating the collisionless Boltzmann equation over
velocity space. In past approximations, such as ZA and its modified
version, velocity dispersion was ignored.
Buchert and Dom\'{\i}nguez~\cite{budo} argued that the effect
of velocity dispersion become important beyond the caustics.
They also argued that models for large-scale structure should
rather be constructed for a flow which describes the average
motion of a multi-stream system. 
Then they showed that when the velocity dispersion is still
small and can be considered isotropic, that gives effective
'pressure' or viscosity terms. Furthermore, they argued the
relation between mass density $\rho$ and pressure $P$, i.e.
an `equation of state'. If the relation between the density of matter
and pressure seems barotropic,
the equation of state should take the form $P \propto \rho^{5/3}$.
Buchert et al.~\cite{bdp} showed how the viscosity term
is generated by the effective pressure of a fluid
under the assumption that the peculiar acceleration is
parallel to the peculiar velocity;
Dom\'{\i}nguez~\cite{domi00,domi0106} clarified that a hydrodynamic
formulation is obtained via a spatial coarse graining
in a many-body gravitating system, and the viscosity term in
the `adhesion approximation' can be derived by the expansion
of coarse-grained equations.
Dom\'{\i}nguez~\cite{domi0103} also reported on a study of the spatially
coarse-grained velocity dispersion in cosmological N-body simulations.
The analysis showed that polytrope exponent becomes $\gamma \simeq 5/3$ in
a quasi-nonlinear regime, and $\gamma \simeq 2$ in a strongly nonlinear
regime. Dom\'{\i}nguez and Melott~\cite{dome0310} discussed polytrope
exponents of velocity dispersion in N-body simulations. According to
their results, the exponents depend on a model of initial density
fluctuation.

From these points, the extension of Lagrangian perturbation
theory to cosmological fluids with pressure has been considered.
Actually, Adler and Buchert~\cite{adler} have formulated the 
Lagrangian perturbation theory for a barotropic fluid.
Morita and Tatekawa~\cite{moritate} and Tatekawa et al.~\cite{tate02}
solved the Lagrangian perturbation equations for a polytropic fluid
up to second order for cases where the equations are solved easily.
Hereafter, we call this model the `pressure model'.

In this paper, we analyze the density field which is described
by the Lagrangian approximations; ZA, TZA,
and first-order pressure model solutions. We calculate the
cross-correlation function of density
field between the Lagrangian approximation and N-body simulation.
Furthermore we analyze the probability distribution function of density
fluctuation for confirmation. From these results,
we determine a polytrope exponent in the equation of state.
 From our analyses
of the cross-correlation coefficient and probability distribution function
of density fluctuation, we find that the value
of the polytrope exponent seems to be $5/3$ for quasi-nonlinear evolution,
as Buchert and Dom\'{\i}nguez argued~\cite{budo}.
However, for the determination of a proportion fixed number
in equation of state,
we must consider further physical processes or carry out high-resolutional
N-body simulation.

This paper is organized as follows.
In Sec.~\ref{sec:Lagrangian}, we present Lagrangian perturbative solutions:
In Sec.~\ref{subsec:first}, we show the first-order solution of pressure
model in the Einstein-de Sitter background.
For comparison,
in Sec.~\ref{subsec:ZA} and \ref{subsec:TZA}, we show the solution of ZA
and the procedure of TZA.

In Sec.~\ref{sec:compari}, we compare the density field between the Lagrangian
approximations and N-body simulation. In Sec.\ref{subsec:CC-function}, we
calculate the cross-correlation coefficient of the density field.
Though it seems
that we can reach a conclusion in this analysis, it is insufficient.
Therefore in Sec.~\ref{subsec:PDF-delta}, we analyze the probability
distribution function of density fluctuation. 
In Sec.~\ref{sec:discuss}, we discuss our results
and state conclusions.

\section{Lagrangian approximations in gravitational instability
theory}\label{sec:Lagrangian}

\subsection{First-order solutions of the pressure model}\label{subsec:first}

In this section, we present perturbative solutions
in the Lagrangian description.
The matter model we consider is a self-gravitating fluid
with mass density $\rho$ and `pressure' $P$,
which is given by the presence of velocity dispersion. 
The `pressure' we adopt here is the same as was introduced
by Buchert and Dom\'{\i}nguez~\cite{budo},
i.e. the diagonal component of the velocity dispersion tensor
when the velocity dispersion is assumed to be isotropic
in the Jeans equation~\cite{BT}.
In Lagrangian hydrodynamics,
the comoving coordinates $\bm{x}$ of the fluid elements are
represented in terms of Lagrangian coordinates $\bm{q}$ as
\begin{equation} \label{x=q+s}
\bm{x} = \bm{q} + \bm{s} (\bm{q},t) \,,
\end{equation}
where $\bm{q}$ are defined as initial values of $\bm{x}$,
and $\bm{s}$ denotes the Lagrangian displacement vector
due to the presence of inhomogeneities.
From the Jacobian of the coordinate transformation from
$\bm{x}$ to $\bm{q}$, $J \equiv \det (\partial x_i / \partial q_j)
= \det (\delta_{ij} + \partial s_i / \partial q_j)$,
the mass density is given exactly as
\begin{equation}\label{exactrho}
\rho = \rho_{\rm b} J^{-1} \,.
\end{equation}

We decompose $\bm{s}$ into the longitudinal
and the transverse modes as
$\bm{s} = \nabla_{\bm{q}} S
+ \bm{S}^{\rm T}$ with
$\nabla_{\bm{q}} \cdot \bm{S}^{\rm T}=0$. In this paper, we show
an explicit form of perturbative solutions in
only Einstein-de Sitter universe.
For a generic background universe, we obtained the perturbation
solutions in our previous paper~\cite{tate02}.

The transverse modes can be solved easily. The first-order solutions
become as follows:
\begin{equation}\label{sol-ST}
\bm{S}^{\rm T} \propto a^0,\; a^{-1/2} \,.
\end{equation}
Because the solutions do not depend on 'pressure', the solutions of
this mode in ZA become the same in form. The transverse modes do not have
a growing solution in a first-order approximation.

For the longitudinal modes, we carry out Fourier transformation
with respect to the Lagrangian coordinates $\bm{q}$.
If we assume a polytropic equation of state $P=\kappa \rho^{\gamma}$
with a constant $\kappa$ and a polytrope exponent $\gamma$,
we can write an explicit form of first-order perturbative solutions.
In the Einstein-de Sitter background,
the solutions are written
in a relatively simple form.
They are, for $\gamma \ne 4/3$,
\begin{equation}\label{hatSbessel}
\widehat{S}(\bm{K},a) \propto a^{-1/4}
\, \mathcal{J}_{\pm 5/(8-6\gamma)}
\left( \sqrt{\frac{2C_2}{C_1}}
\frac{|\bm{K}|}{|4-3\gamma|}
\, a^{(4-3\gamma)/2} \right) \,,
\end{equation}
where $\mathcal{J}_{\nu}$ denotes the Bessel function of order $\nu$,
and for $\gamma=4/3$,
\begin{equation}\label{hatS43}
\widehat{S}(\bm{K},a) \propto
a^{-1/4 \pm \sqrt{25/16 - C_2 |\bm{K}|^2 / 2C_1}} \,,
\end{equation}
where $C_1 \equiv 4 \pi G \rho_{\rm b}(a_{\rm in})
\, a_{\rm in}^{\ 3} /3$
and $C_2 \equiv \kappa \gamma \rho_{\rm b}(a_{\rm in})^{\gamma-1}
\, a_{\rm in}^{\ 3(\gamma-1)}$. $\rho_b$ and $\bm{K}$ mean background
mass density and Lagrangian wavenumber, respectively. $a_{\rm in}$
means scale factor when an initial condition was given.

Here we notice the relation between the behaviors of the above solutions
and the Jeans wavenumber, which is defined as
\[
K_{\rm J} \equiv \left(
\frac{4\pi G\rho_{\rm b} a^2}
     {{\rm d} P / {\rm d} \rho (\rho_{\rm b})} \right)^{1/2} \,.
\]
The Jeans wavenumber, which gives a criterion for
whether a density perturbation with a wavenumber
will grow or decay with oscillation,
depends on time in general.
If the polytropic equation of state $P=\kappa\rho^{\gamma}$
is assumed,
\begin{equation}\label{kjeans}
K_{\rm J} = \sqrt{\frac{3C_1}{C_2}} \, a^{(3\gamma-4)/2} \,.
\end{equation}
Equation~(\ref{kjeans}) implies that, if $\gamma < 4/3$,
$K_{\rm J}$ will be decreased and
density perturbations with any scale will decay and disappear
in evolution,
and if $\gamma > 4/3$, all density perturbations will
grow to collapse.
We rewrite the first-order solution Equation~(\ref{hatSbessel}) with
the Jeans wavenumber:
\begin{equation}
\widehat{S}(\bm{K},a) \propto a^{-1/4}
\, \mathcal{J}_{\pm 5/(8-6\gamma)}
\left( \frac{\sqrt{6}}{|4-3\gamma|}
\frac{|\bm{K}|}{K_{\rm J}} \right) \,.
\end{equation}
%

\subsection{Zel'dovich approximation}\label{subsec:ZA}

ZA was obtained as first-order
solutions with dust fluid in Lagrangian description~\cite{zel}. The
solutions are obtained from the solutions of the pressure model as a
limit of weak pressure. For example, in E-dS model, when
we take a limit $\kappa \rightarrow0$ in Equations
(\ref{hatSbessel}) and (\ref{hatS43}),
the solutions converge to those of ZA:
\begin{equation}
\widehat{S}(\bm{K},a) \propto a,\; a^{-3/2} \,.
\end{equation}

ZA is known as perturbative solutions which describe the structure
well in quasi-linear scale. However if the caustics appear, the solutions
no longer have physical meaning.

\subsection{Truncated Zel'dovich approximation}\label{subsec:TZA}

During
evolution, the small scale structure contracts and forms caustics. Therefore
if we introduce some cutoff in the small scale, we will be able to avoid 
the formation of caustics~\cite{cms,mps}. In TZA, for the avoidance
of caustics, we introduce a Gaussian cutoff to the initial density
spectrum as follows:
\begin{equation}
{\cal P}(k,t_{\rm in}) \rightarrow {\cal P}(k, t_{\rm in}) \exp
\left (-k^2/k_{NL} \right ) \,,
\end{equation}
where $k_{NL}$ means `nonlinear wavenumber', defined by
\begin{equation} \label{eqn:TZA-knl}
1 = a(t)^2 \int_{k_0}^{k_{NL}} {\cal P}(k,t_{\rm in}) dk \,.
\end{equation}
The `nonlinear wavenumber' depends on the scale factor.
The relation between the Jeans wavenumber $K_J$ and the nonlinear
wavenumber $k_{NL}$ will be discussed in Sec.~\ref{sec:discuss}.

\section{Comparison between N-body simulation and Lagrangian approximations}\label{sec:compari}

In this section, we show a comparison between N-body simulation and
Lagrangian approximations with two statistical quantities.
In our previous paper~\cite{tate02}, we showed that the effect
of second order perturbation was still small just before shell-crossing.
Therefore we consider only first order perturbation.

We analyze ZA~\cite{zel}, TZA~\cite{cms,mps}, and the pressure
model~\cite{adler,moritate,tate02}. We establish the value of scale
factor at $z=0$ with $a=1$.
For the initial condition, we set the Gaussian density field with
scale-free spectrum;
\begin{equation}
\mathcal{P}(k) \propto k^n (n=-1, 0, 1) \,,
\end{equation}
SCDM, and LCDM model. The initial condition was produced
by COSMICS~\cite{COSMICS}.

For N-body simulation, we execute $P^3 M$ code. The parameters
of simulation were given as follows:
\begin{eqnarray*}
\mbox{Number of particles} &:& N=64^3, N=128^3
 \mbox{(Figure~\ref{fig:Cor-k1-N128} only)} \,, \\
\mbox{Box size} &:& L=64 h^{-1} \mbox{Mpc} \,, \\
\mbox{Softening Length} &:& \varepsilon= 0.05 h^{-1} \mbox{Mpc} \,, \\
\mbox{Course-Graining Length} &:& l = 1, 2, 4 h^{-1} \mbox{Mpc} \,, \\
\mbox{Hubble patameter} &:& h=0.71 \,.
\end{eqnarray*}

For CDM models, we choose cosmological parameters as follows:
\begin{eqnarray*}
\mbox{SCDM} &:& \Omega_m=1.0,\; \Omega_{\Lambda}=0.0,\; \sigma_8=0.84 \\
\mbox{LCDM} &:& \Omega_m=0.27,\; \Omega_{\Lambda}=0.73,\; \sigma_8=0.84 
\end{eqnarray*}

In the pressure model, we choose a polytrope exponent $\gamma=4/3, 5/3$:
In the case of $\gamma=4/3$, we obtain the simplest perturbative solution
given by Equation~(\ref{hatS43})
that is described by the power-law of the time variable.
$\gamma=5/3$ is given from theoretical argumentation by
Buchert and Dom\'{\i}nguez~\cite{budo}. They argued kinematic considerations
using collisionless Boltzmann equation and derived $\gamma=5/3$. The 
exponent is equivalent to the adiabatic process of an ideal gas. Because
we cannot decide on a proportional constant $\kappa$ in the equation of state
from past discussion, we choose several values. In this paper, instead
of $\kappa$, we write an initial ($a=10^{-3}$, i.e. $z=1000$) Jeans wavenumber,
given by Equation~(\ref{kjeans}).

Here we show how we set up the initial condition in the pressure model.
We adjust the initial condition in the pressure model to be the same
as that in ZA: the initial peculiar velocity in the pressure model is
same as that in ZA.
The procedure for setting up initial condition was shown in our
previous papers~\cite{moritate,tate02}.%
%

\subsection{Cross-correlation coefficient}\label{subsec:CC-function}

First we calculate the cross-correlation coefficient of density fields.
The Cross-correlation coefficient was used for the comparison
of the resulting density fields
~\cite{cms,mps,msw,Buchert94,Melott95,Karakatsanis97}.
The cross-correlation coefficient is defined by
\begin{equation}
S \equiv \left< \frac{\delta_1 \delta_2}
 {\sigma_1 \sigma_2} \right> \,,
\end{equation}
where $\sigma_i$ means the density dispersion of model $i$,
\begin{equation}
\sigma_i \equiv \sqrt{\left < \delta_i^2 \right >} .
\end{equation}
$S=1$ means that the pattern of density field of two models
coincide with each other. In linear regime, the density dispersion
remains $\sigma \ll 1$. Although we develop the structure until
it becomes
strongly nonlinear regime ($\sigma_{N-body} > 1$), we especially
analyze it in the quasi-nonlinear regime ($\sigma_{N-body} \simeq 1$).

\begin{figure}
 \includegraphics{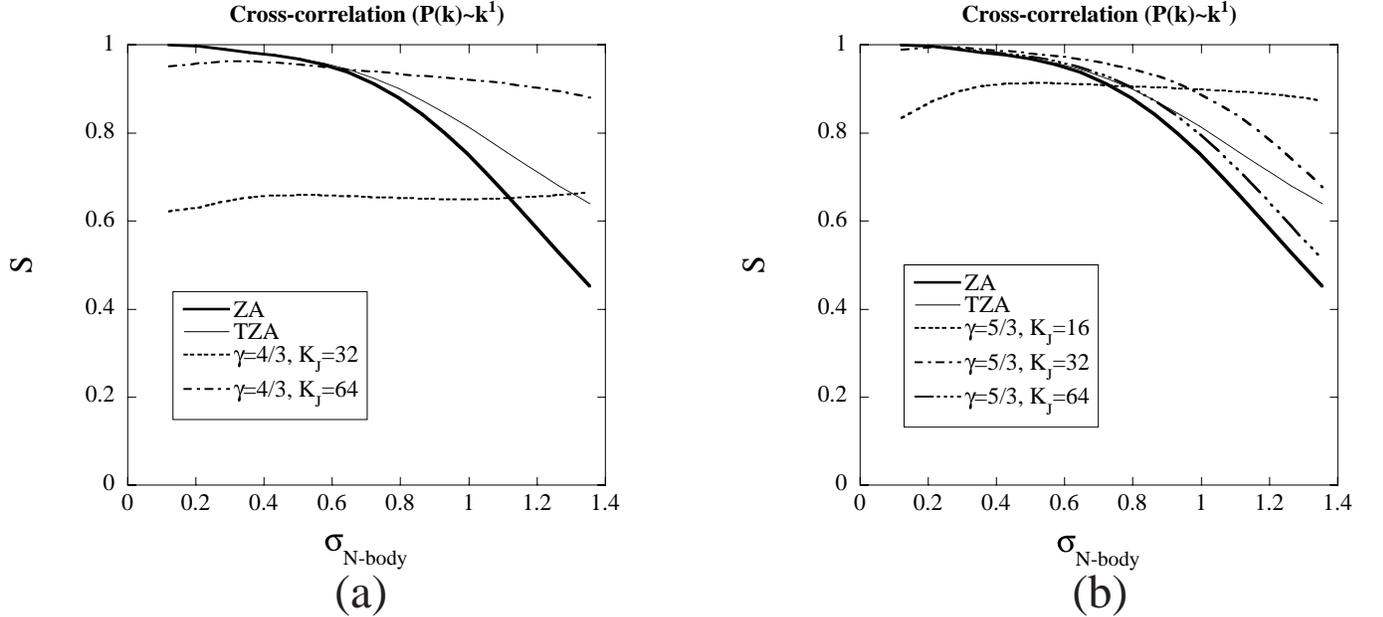}
 \caption{The cross-correlation coefficient of density fields
 between the N-body simulation and Lagrangian approximations.
 Primordial density fluctuation is given by the scale-free spectrum
 $P(k) \propto k^1$ ($N=64^3, l = 1 h^{-1}
 \mbox{Mpc}$).
 ~(a) When we choose $\gamma=4/3$, the function
 deviates from that of ZA in linear regime.~(b) In the case of
 $\gamma=5/3$, we can obtain a better result than TZA.}
\label{fig:Cor-k1}
\end{figure}

\begin{figure}
 \includegraphics{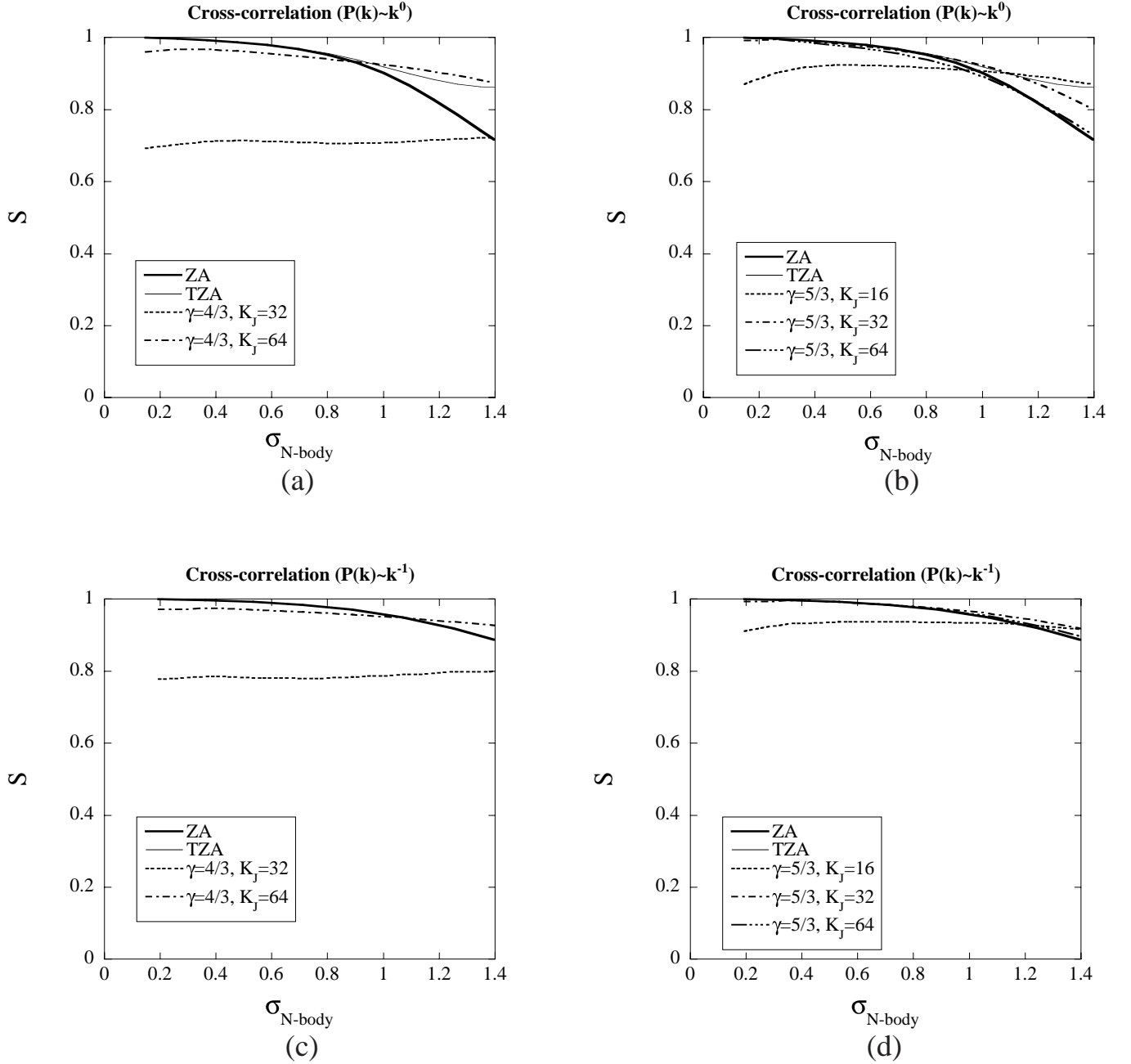}
 \caption{The cross-correlation coefficient of density fields
 between the N-body simulation and Lagrangian approximations
 ($N=64^3, l = 1 h^{-1} \mbox{Mpc}$, scale-free spectrum model).
 ~(a) $P(k) \propto k^0$. The case of
 $\gamma=4/3$. ~(b) $P(k) \propto k^0$, the case of $\gamma=5/3$.
 ~(c) $P(k) \propto k^{-1}$. In this model, the difference in
 the coefficient between ZA and TZA becomes small. The case of
 $\gamma=4/3$. ~(d) $P(k) \propto k^{-1}$, the case of $\gamma=5/3$.}
\label{fig:Cor-k0}
\end{figure}

\begin{figure}
 \includegraphics{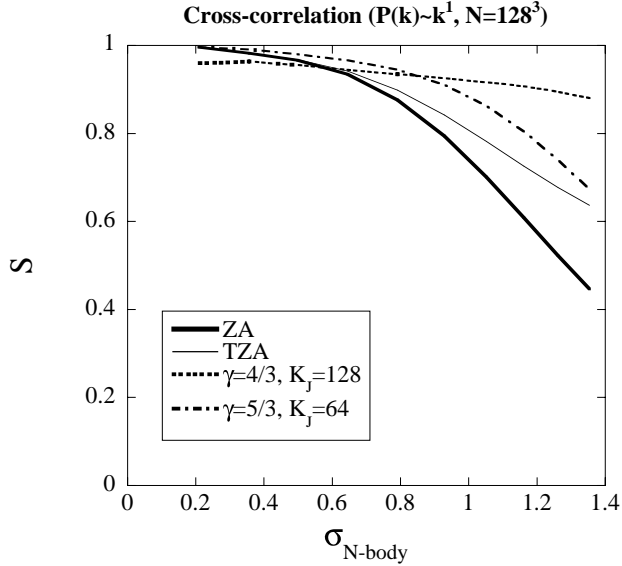}
 \caption{The cross-correlation coefficient of density fields
 between N-body simulation and Lagrangian approximations.
 Primordial density fluctuation is given by scale-free spectrum
 $P(k) \propto k^1$ ($N=128^3, l = 1 h^{-1}
 \mbox{Mpc}$). From comparison between Figure~\ref{fig:Cor-k1} and
 this graph, we can notice that the results depend on the ratio
 of grid size and initial Jeans wavenumber.}
\label{fig:Cor-k1-N128}
\end{figure}

\begin{figure}
 \includegraphics{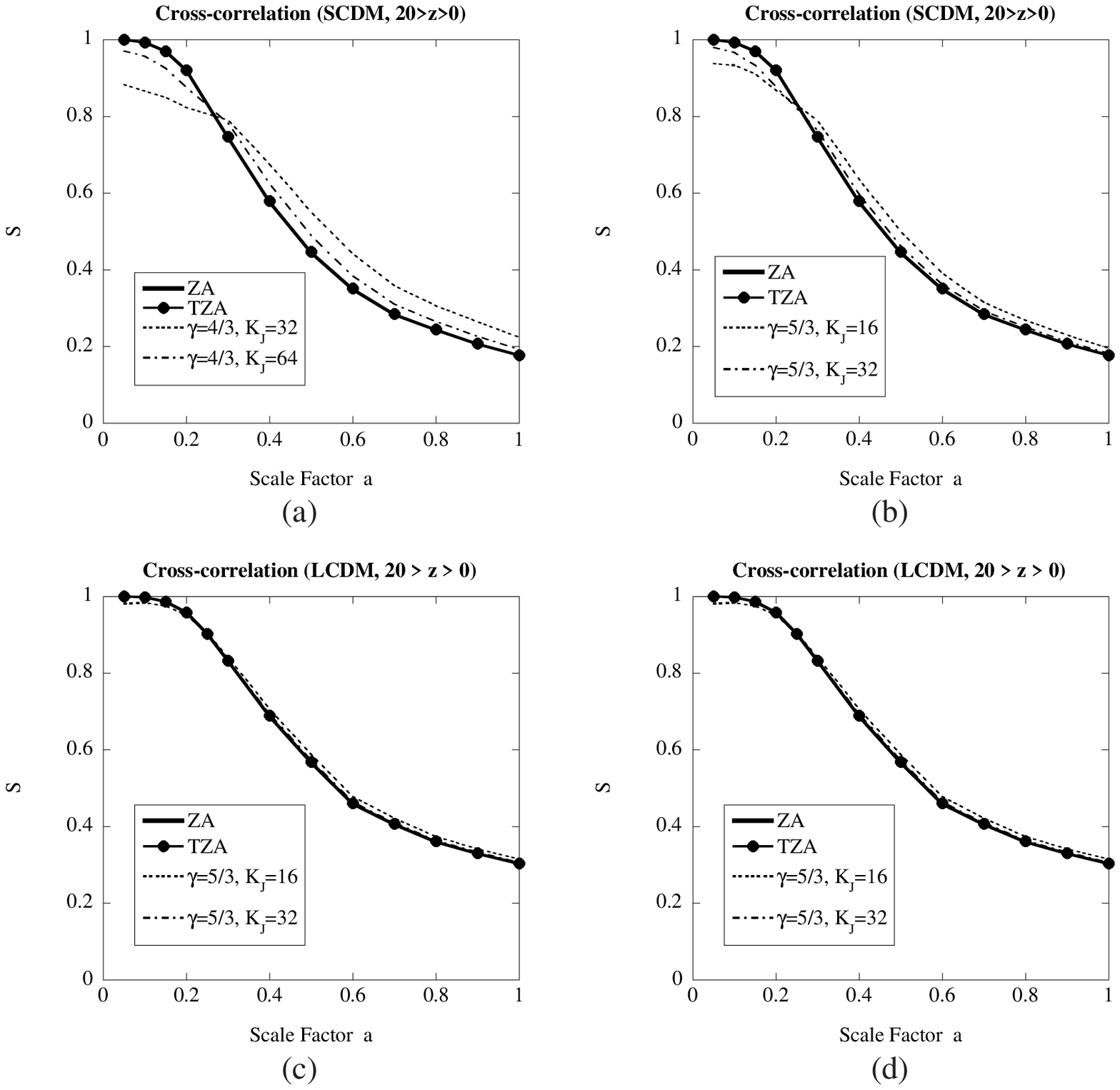}
 \caption{The cross-correlation coefficient of density fields
 between an N-body simulation and Lagrangian approximations.
 The primordial density fluctuation is given by the CDM spectrum
 ($N=64^3, l = 1 h^{-1} \mbox{Mpc}$). The model in
 which hardly a difference appears in is excluded from the graph.
 ~(a) The SCDM model with $\gamma=4/3$.
 ~(b) The SCDM model with $\gamma=5/3$.
 ~(c) The LCDM model with $\gamma=4/3$.
 ~(d) The LCDM model with $\gamma=5/3$.}
\label{fig:Cor-CDM}
\end{figure}

\begin{figure}
 \includegraphics{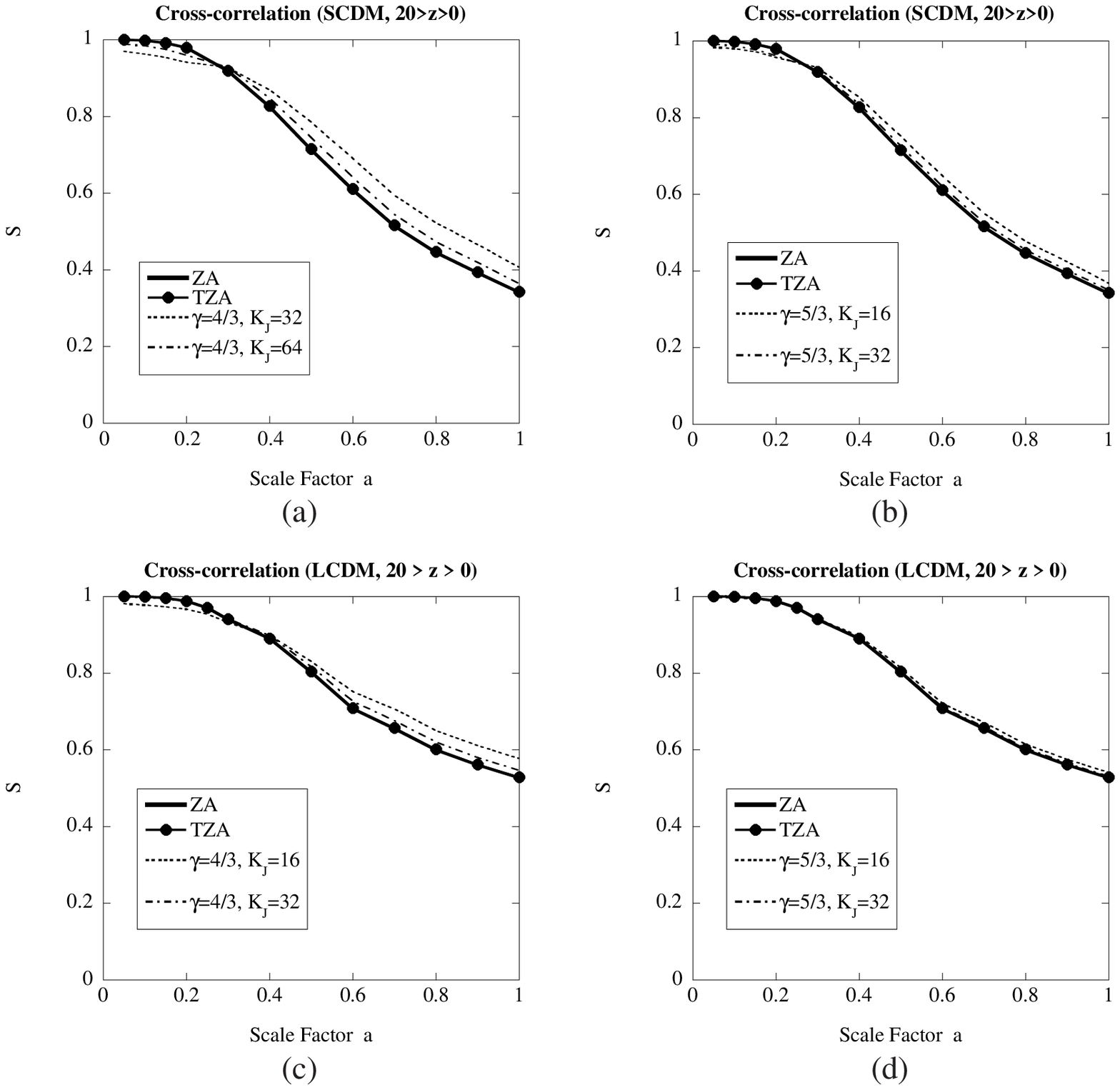}
 \caption{The same as Figure~\ref{fig:Cor-CDM}.
 In these figures, we changed the course-grained length to
 $l = 2 h^{-1} \mbox{Mpc}$.
 ~(a) The SCDM model with $\gamma=4/3$.
 ~(b) The SCDM model with $\gamma=5/3$.
 ~(c) The LCDM model with $\gamma=4/3$.
 ~(d) The LCDM model with $\gamma=5/3$.}
\label{fig:Cor-CDM2}
\end{figure}

\begin{figure}
 \includegraphics{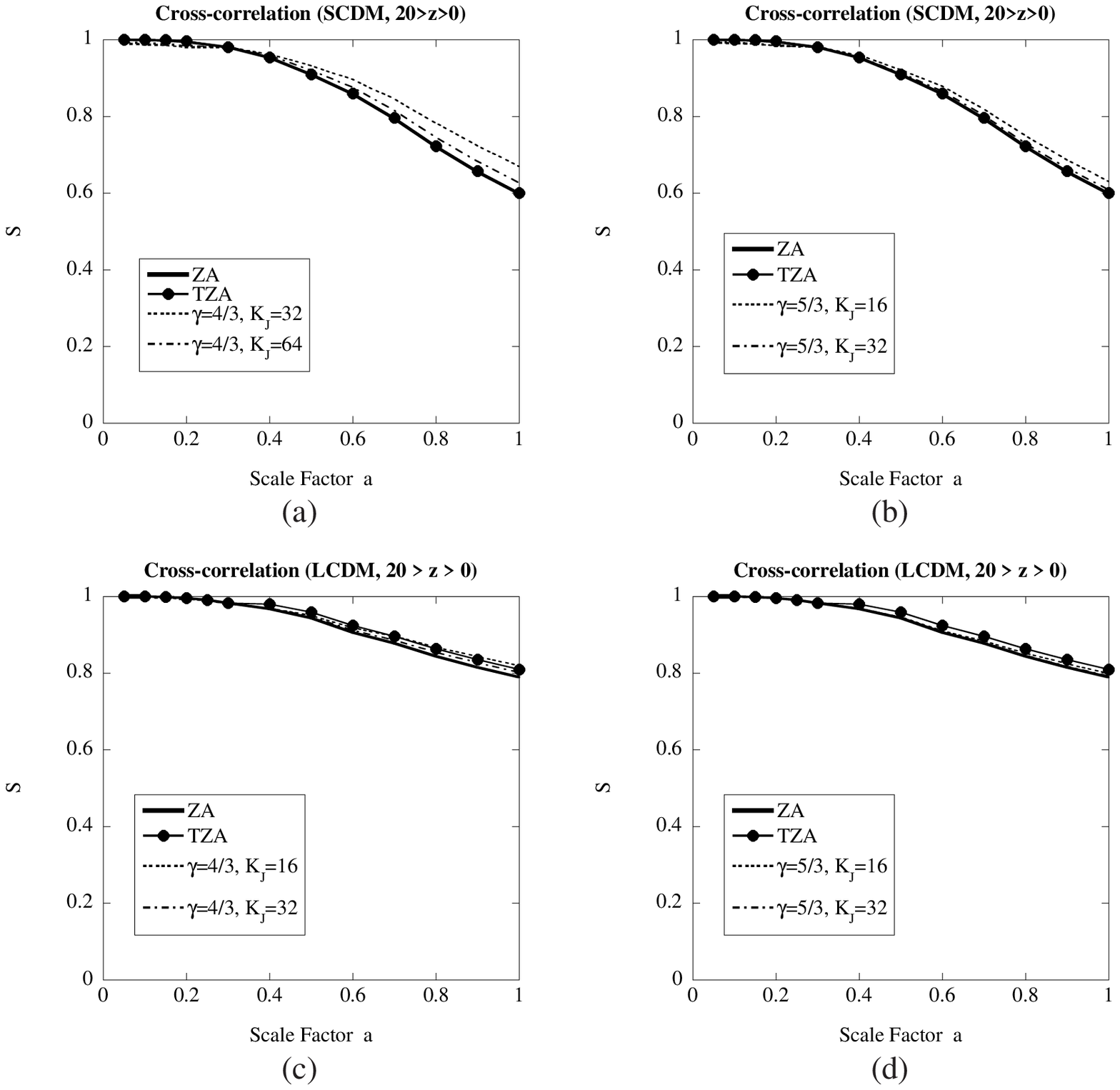}
 \caption{The same as Figure~\ref{fig:Cor-CDM}. In these figures,
  we changed the course-grained length to $l = 4 h^{-1} \mbox{Mpc}$.
 ~(a) The SCDM model with $\gamma=4/3$.
 ~(b) The SCDM model with $\gamma=5/3$.
 ~(c) The LCDM model with $\gamma=4/3$.
 ~(d) The LCDM model with $\gamma=5/3$.}
\label{fig:Cor-CDM4}
\end{figure}

Figure~\ref{fig:Cor-k1}-\ref{fig:Cor-CDM4} shows a comparison of N-body
density fields with those predicted by various Lagrangian approximations.
First, we notice scale-free spectrum cases
(Figure~\ref{fig:Cor-k1}-\ref{fig:Cor-k1-N128}). As in the past
analyses, TZA shows better performance than ZA. Our analyses also
show similar tendency, i.e. our analyses do not contradict past
analyses.

In the pressure model, the
performance strongly depends on the polytrope exponent $\gamma$ and
 Jeans wavenumber.
In the case of $\gamma=4/3$, when we set the initial Jeans wavenumber
to be small, even if in linear regime, the approximation deviates from
an N-body simulation. Only for the case of $K_J=64$ the approximation shows
better performance than ZA in a quasi-nonlinear regime. We notice that
the result strongly depends on the Jeans wavenumber in the case of
$\gamma=4/3$:
When we slightly change the value of the Jeans wavenumber, the
cross-correlation coefficient changes dramatically. In the case of
$\gamma=5/3$, although the result depends slightly on
initial Jeans wavenumber, the pressure model shows a better performance
than ZA in a quasi-nonlinear regime. Furthermore, the pressure model
also shows a better performance than TZA. However, when we consider
scale-free spectrum models, the model does not have typical
physical scale: the model
has only box size, grid size, and softening length. The trend of the
result was unchanged when we changed the box size of the model
and the softening parameter. When we change the number of particles, the
result changes. From a comparison of Figure~\ref{fig:Cor-k1} and
Figure~\ref{fig:Cor-k1-N128}, we can notice that the results depend on
the ratio of grid size and initial Jeans wavenumber.
In our calculation, we found out that it was good to set
up  the value of $\kappa$ so that the initial ($z=1000$) Jeans wavenumber
$K_J$ was
$N^{1/3}/4 \le K_J \le N^{1/3}$. For example, in the case of $N=64^3$,
as we see
in Figure~\ref{fig:Cor-k1} and \ref{fig:Cor-k0}, it is good to choose
the initial Jeans wavenumber $16 \le K_J \le 64$.

Next we consider SCDM and LCDM models (Figure~\ref{fig:Cor-CDM}-
\ref{fig:Cor-CDM4}).
In these models, the difference between
ZA and TZA becomes very small. Because the initial density spectrum in
the CDM model dumps power in a small scale, the cutoff in spectrum weakly
affects the formation of caustics, as we saw in the case of
$P(k) \propto k^{-1}$.
From Figure~\ref{fig:Cor-CDM}, we can see that
the effect of pressure improves the approximation
in the quasi-nonlinear stage.
Also, in both the SCDM and LCDM models, the case of $\gamma=4/3$
shows deviation
from ZA in linear regime. On the other hand, the case of $\gamma=5/3$ shows
that the cross-correlation coefficient becomes almost the same
in linear regime.
In the case of $\gamma=4/3$, when we choose a small initial
Jeans wavenumber
(for example, $K_J=16$),
although we can improve the approximation much more in quasi-nonlinear stage
than in large Jeans wavenumber ($K_J=32, 64$) cases, the approximation
changes slightly for the worse in the linear stage. On the other hand,
when we choose $\gamma=5/3$, although the effect seems small,
we can obtain a improved solution both in linear and in
quasi-nonlinear stages.

When the model evolves to strongly nonlinear regime, the trend of
the solutions change. In the linear stage, the case of $\gamma=4/3$
shows deviation from ZA.
However in strongly nonlinear regime, though the Lagrangian approximation
generally become worse, the case of $\gamma=4/3$ shows a rather good result
(Figure~\ref{fig:Cor-CDM} (a) and (c)).
This tendency was unchanged even though the course-grained
length was changed
(Figure~\ref{fig:Cor-CDM2} and \ref{fig:Cor-CDM4}).

In both the SCDM and LCDM models, when we choose a small
initial Jeans wavenumber $K_J$, although the approximation is improved
after the quasi-nonlinear stage, the reasonable range of Jeans wavenumbers
seems wide.
The strict limitation to the value of $\kappa$ or the initial Jeans
wavenumber
will be given by other physical considerations or by the
high-resolution N-body simulation.

From these results, we find that it is reasonable to choose the polytrope
exponent $\gamma=5/3$ until quasi-nonlinear regime is reached.
These results support the
suggestion by Buchert and Dom\'{\i}nguez~\cite{budo} and by
Dom\'{\i}nguez~\cite{domi0103}.
If we have interest in strongly nonlinear regime, although
the Lagrangian
approximation generally becomes worse, we had better analyze the case of
$\gamma=4/3$. In this regime, it is necessary to consider whether that
approximation will still be valid. In any case, from the cross-correlation
coefficient, we can give limitation in the polytrope exponent $\gamma$.

Unfortunately, in these results, we cannot give a strict limit to the
proportion coefficient $\kappa$ of the equation of state.
When we choose $\gamma=4/3$,
we show that the result strongly depends on $\kappa$, and we can notice
a strict limitation. However when we choose $\gamma=5/3$,
we can hardly judge the best value for
$\kappa$. In our calculation, we found out that it was good to set up 
the initial ($a=10^{-3}$, i.e. $z=1000$)
Jeans wavenumber $K_J$ as
$N^{1/3}/4 \le K_J \le N^{1/3}$.
From the range of $K_J$, we can obtain a reasonable value for $\kappa$.
If we choose large value for $\kappa$,
it becomes hard to form nonlinear structure. On the other hand,
if we choose small value for $\kappa$, the structure becomes almost
the same as the structure
which was obtained by ZA.

Although the cross-correlation coefficient is one thing which is
good for checking the accuracy of the approximation,
it is not enough for checking. Now we consider two
samples A and B. The density contrast of the samples is given by
$\delta_A$ and $\delta_B$, respectively. We assume that the following
proportion relation exists
between $\delta_A$ and $\delta_B$:
\begin{equation}
\delta_A \propto \delta_B \,.
\end{equation}
Even if the density contrast is greatly different, as $\delta_A$ and $\delta_B$,
and one shows a highly nonlinear structure and another remains
in the linear regime,
the cross-correlation coefficient between $\delta_A$ and $\delta_B$ becomes $1$. 

Therefore, we must check the accuracy of the approximation with
another property.
In the next subsection, we analyze the probability distribution function
of density fluctuation.

\subsection{Probability Distribution Function of Density Fluctuation}
\label{subsec:PDF-delta}

\begin{figure}
 \includegraphics{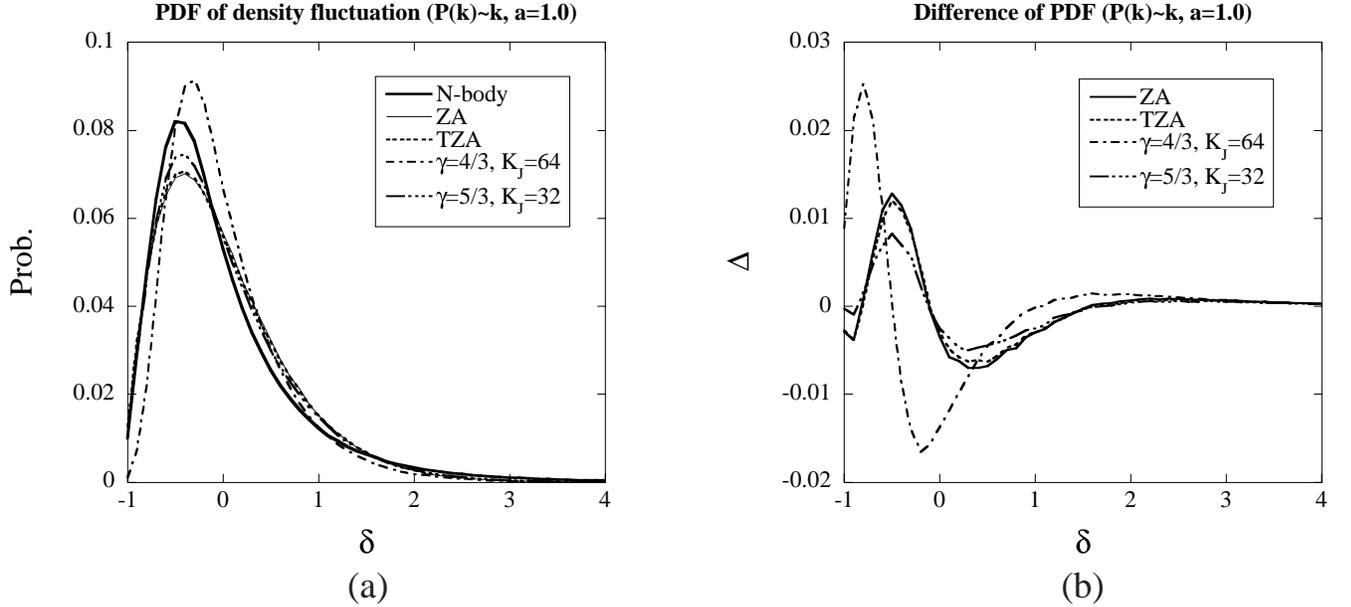}
 \caption{The PDF of density fluctuation for a scale-free spectrum
 ($P(k) \propto k$, $l=1 h^{-1} \mbox{Mpc}$: $\sigma_{N-body} \simeq 1$
  at $a=1.0$.).
 ~(a) The PDF of density fluctuation. In the case of $\gamma=4/3$, the effect of
  pressure suppresses the growth of the fluctuation.
 ~(b) The difference in the PDF of density fluctuation. In this figure,
 the difference between the case
 of $\gamma=4/3$ and other cases becomes clear. When we choose $K_J=32$
 for the case of $\gamma=4/3$, more greater difference appears.
 }
\label{fig:PDF-k1}
\end{figure}

\begin{figure}
 \includegraphics{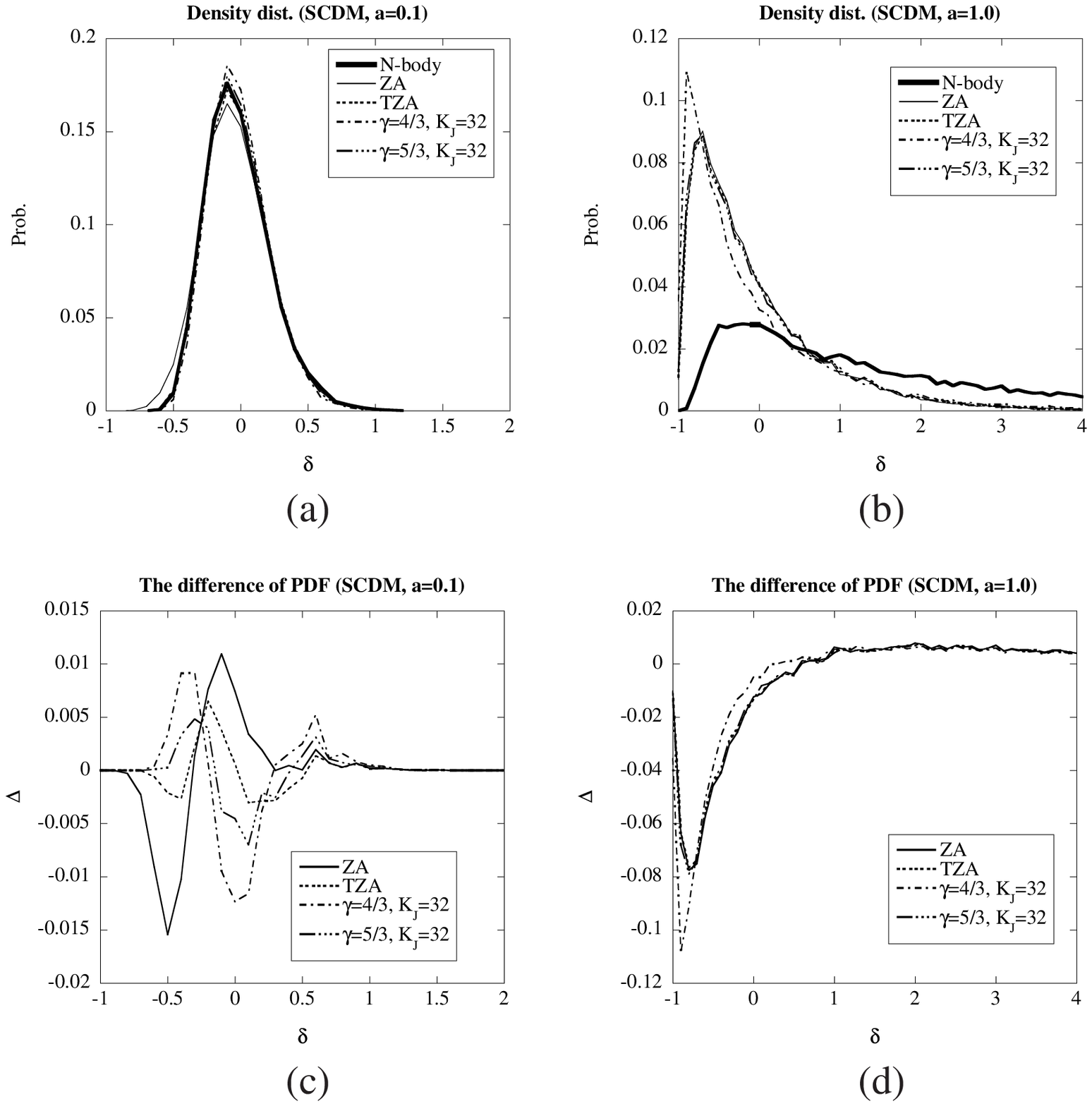}
 \caption{The PDF of density fluctuation in the SCDM model
 ($l=4 h^{-1} \mbox{Mpc}$).
 In the case of $\gamma=4/3$, the pressure effect suppresses the growth of
 density fluctuation. Therefore the probability of a small fluctuation
 ($|\delta|<1$) increases.
 ~(a) The SCDM model, at $a=0.1$ ($z=9$, $l=4 h^{-1} \mbox{Mpc}$.
  Quasi-nonlinear regime). In the case of $\gamma=4/3$, the effect of
  pressure suppresses the growth of the fluctuation.
 ~(b) The SCDM model, at $a=1.0$ ($z=0$, $l=4 h^{-1} \mbox{Mpc}$.
  Strongly nonlinear regime).
 ~(c) The difference in the PDF of density fluctuation between
  the N-body simulation and Lagrangian approximations. At $a=0.1$.
 ~(d) Same as (c), but at $a=1.0$.}
\label{fig:PDF-SCDM}
\end{figure}

Here, we compare the probability distribution function (PDF) of
density fluctuation.
In the Eulerian linear approximation, if initial data is given by a random
Gaussian distribution, the PDF of density fluctuations will retain its
Gaussianity during evolution. On the other hand, in the Lagrangian
approximation, there appears a nonlinear effect. In fact,
Kofman et al.~\cite{Kofman94} shows that the PDF of density fluctuation
approaches a log-normal function
rather than a Gaussian function in the cases of the Lagrangian
approximation and N-body simulation. Padmanabhan and
Subramanian~\cite{PadSub93} also discussed the PDF of density fluctuation
with the ZA and found
a non-Gaussian distribution.

\begin{figure}
 \includegraphics{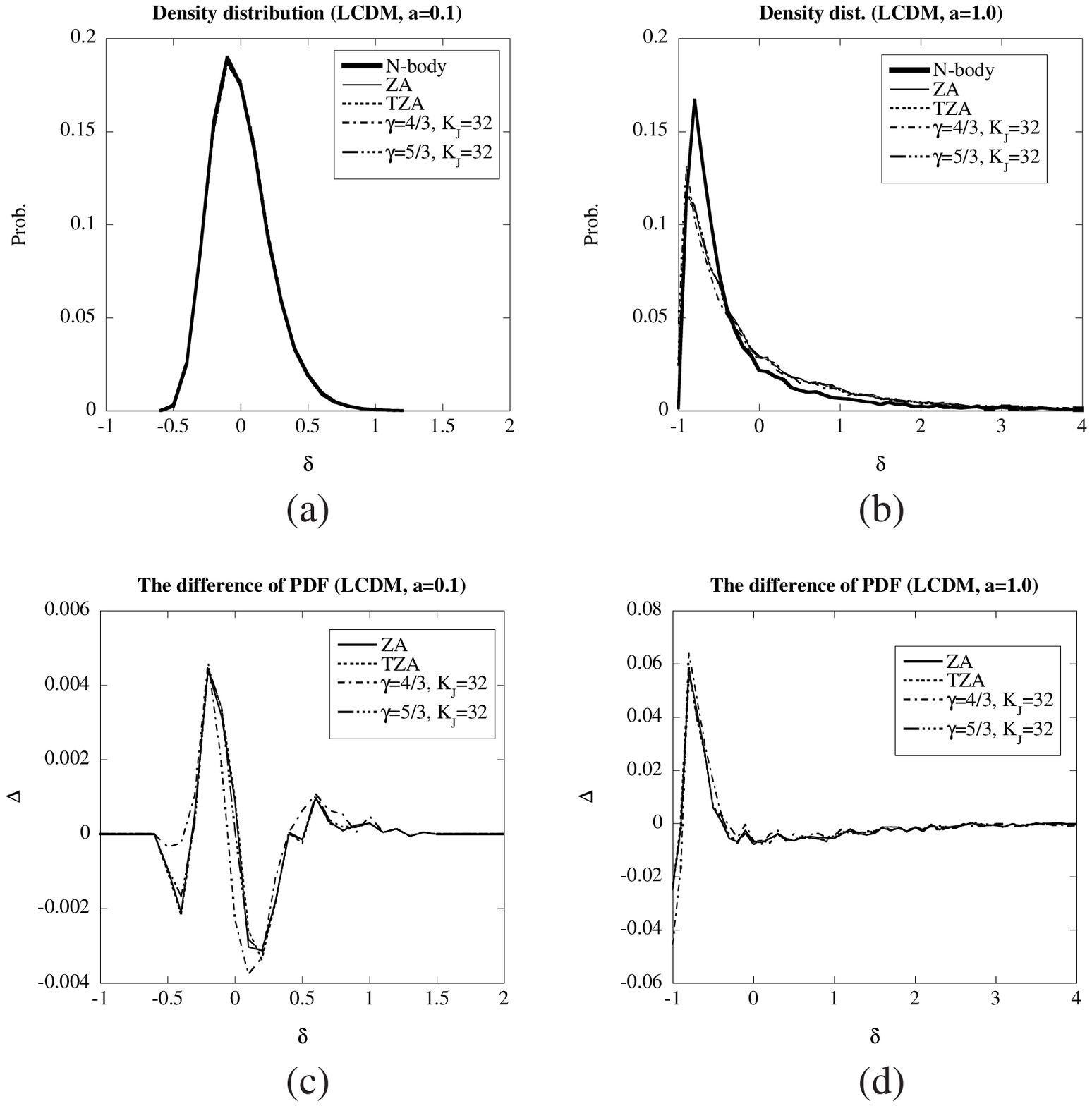}
 \caption{The PDF of density fluctuation in the LCDM model
 ($l=4 h^{-1} \mbox{Mpc}$).
 In the case of $\gamma=4/3$, the pressure effect suppresses the growth of
 density fluctuation. Therefore the probability of a small fluctuation
 ($|\delta|<1$) increases.
 ~(a) At $a=0.1$ ($z=9$, $l=4 h^{-1} \mbox{Mpc}$.
  Quasi-nonlinear regime). The PDFs of density fluctuation
  seem similar to each other.
 ~(b) At $a=1.0$ ($z=0$, $l=4 h^{-1} \mbox{Mpc}$.
  Strongly nonlinear regime).
 ~(c) The difference in the PDF of density fluctuation between
  the N-body simulation
  and Lagrangian approximations. At $a=0.1$.
 ~(d) Same as (c), but at $a=1.0$.}
\label{fig:PDF-LCDM}
\end{figure}

How will the PDF of density fluctuation change if we take the effect of
the velocity dispersion into consideration?
Figures~\ref{fig:PDF-k1}, \ref{fig:PDF-SCDM}, and \ref{fig:PDF-LCDM}
shows the PDF
of density fluctuation. As in past works, the PDF of density fluctuation
becomes log-normal in form in the N-body simulation.
In Figures~\ref{fig:PDF-k1}, \ref{fig:PDF-SCDM}, and \ref{fig:PDF-LCDM},
the cases of $\gamma=4/3$ obviously show
the different tendency: in these cases, the effect of pressure suppress
the growth of positive fluctuation
 (Figure~\ref{fig:PDF-k1}(b), \ref{fig:PDF-SCDM} (c),(d),
 and \ref{fig:PDF-LCDM} (c),(d)).
When we also consider the PDF of density fluctuation,
we can notice that it is not so good to choose $\gamma=4/3$
to examine the growth of structure, though the cross-correlation
coefficients well show the trend. On the other hand, the case of  $\gamma=5/3$
well show the trend in the
PDF of density fluctuation.  Although the difference of distribution
between ZA and the pressure model is still small in quasi-nonlinear regime,
the
effect of the pressure can promote the evolution of nonlinear structure.
Therefore the probability of low and high density regions increases in
the case of $\gamma=5/3$.
Furthermore, according to Figure~\ref{fig:PDF-SCDM}(c),
the PDF of density fluctuation in the cases of $\gamma=5/3$ show
that it is much better than that in the TZA case.
Of course when we reach a strongly nonlinear regime, it is necessary
to consider whether that approximation is still valid or not.

From both the cross-correlation coefficient and PDF of density fluctuation,
we can decide that it is reasonable to choose $\gamma=5/3$ as
the polytrope exponent of the equation of state. However it is hard to decide
the proportional parameter $\kappa$. From the results in this paper, we
cannot give tight limit to $\kappa$. To decide the value of $\kappa$, we
will analyze a high-resolution N-body simulation or consider other
physical processes.
For example, we will consider the effect of anisotropic velocity
dispersion~\cite{mtm} or the higher order velocity cumulant.

\section{Discussion and Concluding Remarks}
\label{sec:discuss}

We compared two statistical quantities between an N-body simulation
and Lagrangian approximations. In our
earlier work~\cite{moritate,tate02}, we solved the first-order perturbation
equations in the homogeneous and isotropic background, and the second-
order ones explicitly for the case $\gamma =4/3, 5/3$ in Einstein-de
Sitter Universe. We showed that
the difference between the Lagrangian first-order and second-order
approximations becomes small in the case of $\gamma \ge 4/3$. Therefore,
in this paper we consider only the first-order perturbative solution
for the case $\gamma=4/3, 5/3$. Then we carried out similar calculation
with ZA and TZA to examine their difference from the past models.

First, we compared these models by the cross-correlation coefficient
of the density field between the N-body simulation and Lagrangian
approximations. In
scale-free spectrum cases, as well as in the previous
analyses, TZA shows a better performance than ZA. In the pressure model, the
performance strongly depends on polytrope exponent $\gamma$ and Jeans wavenumber.
In the case of $\gamma=4/3$, when we set that initial Jeans wavenumber
to be small, even in linear regime the approximation deviates from
the N-body simulation. In the case of $\gamma=5/3$, although the result
slightly depends on the initial Jeans wavenumber, the pressure model
shows a better performance than ZA in quasi-nonlinear regime.
Furthermore, the pressure model also shows better performance than TZA.
In the SCDM and LCDM models, the case of $\gamma=4/3$ shows deviation
from ZA in linear regime. On the other hand, the case of $\gamma=5/3$ shows
that the cross-correlation coefficient becomes almost the same
in linear regime.
When the model reaches a strongly nonlinear stage, although the Lagrangian
approximation
generally become worse, the case of $\gamma=4/3$ shows a rather good result.
Of course, in this regime, it is necessary to consider
whether that approximation is still valid.

Second, we analyzed the PDF of density fluctuation.
The cases of $\gamma=4/3$ obviously show a different tendency until
quasi-nonlinear regime is reached: in this case, the effect of
pressure suppresses the
growth of structure. When we also consider the probability distribution of
density, we can see that it is not so good to choose $\gamma=4/3$
to examine the growth of structure, although the cross-correlation
coefficients perform
well. On the other hand, the case $\gamma=5/3$ shows good tendency
in the PDF of density fluctuation.  Although the difference of
the PDF of density fluctuation
between ZA and the pressure model is still small in quasi-nonlinear
regime, the effect of the pressure can promote the evolution of
nonlinear structure. 
The difference between the models of Lagrangian approximation becomes
small when we calculate the evolution until strongly nonlinear regime
is reached.
From analyses of the cross-correlation coefficient
of density field and the PDF
of density fluctuation, we can decide that it is reasonable to
choose $\gamma=5/3$ as the polytrope exponent of the equation of state.

In this paper, we changed some values of the Jeans wavenumber
$K_J$ and undertook analysis. Will there be any relations between
the `nonlinear wavenumber' $k_{NL}$ in TZA
and $K_J$ ? The correspondence is as follows. For simplification,
we consider the correspondence in case of a scale-free spectrum\
$P(k) \propto k^n$.
According to the definition of a `nonlinear wavenumber'
in TZA, $k_{NL}$ is given from Equation~(\ref{eqn:TZA-knl}).
In case of the scale-free spectrum
$P(k) = A k^n$, the definition becomes
\begin{equation}
\frac{1}{n+1} a(t)^2 A k_{NL}^{n+1} =1 .
\end{equation}
From this definition, $k_{NL}$ is written as
\begin{equation}
k_{NL} \sim a^{-2/(1+n)} .
\end{equation}
For example, when we choose $n=1$, $k_{NL}$ becomes
\begin{equation}
k_{NL} \sim \frac{1}{a}.
\end{equation}
On the other hand, the Jeans wavenumber $K_J$ in the pressure model
is given from Equation~(\ref{kjeans}). When we choose $\gamma=2$, $K_J$ becomes
\begin{equation}
K_J \sim \frac{1}{a}.
\end{equation}

There are some different points to consider when we think about time
evolution, although the relation seems to be as described above: First,
in TZA, $k_{NL}$ affects only the initial spectrum. On the other hand,
$K_J$ affects the evolution of fluctuation. Second, although
$k_{NL}$ obviously depends on the initial spectrum, we did not clarify
the dependence on the initial
condition of $K_J$. We think that a consideration of the physical
process, which was not considered here, or the analysis of
the N-body simulation, is necessary for a decision of $K_J$,
i.e. $\kappa$.
We will have to think about the correspondence between the adhesion
approximation and the pressure model. Buchert et al.~\cite{bdp}
showed how the viscosity term in the adhesion approximation is
generated by a pressure-like force.
Dom\'{\i}nguez~\cite{domi00, domi0106} discussed spatial coarse
graining in a gravitating system and derived an evolution equation of
`adhesion approximation'. In the pressure model, we showed that the
density distribution of the pressure model was similar to that
of TZA in the previous paper~\cite{tate02}. The acute characteristic
skeleton structure which appeared in the adhesion approximation
could not be seen from the calculations in our previous paper.
We will consider the relation between the viscosity term in
coarse-grained equations and the pressure term in our model. Then
we will analyze the correspondence between the viscosity term in
the adhesion approximation and the proportional constant $\kappa$
in equation of state in pressure model.

In this paper, we analyzed only density distribution. How will peculiar
velocity distribution change with the effect of 'pressure'? In ZA, the
peculiar velocity is in proportion to the Lagrangian displacement. Then
the growth rate of perturbation is independent of scale. Therefore,
although the structure becomes nonlinear regime, if the initial
condition is 
given as Gaussian, the peculiar velocity distribution remains
Gaussian all the time~\cite{Kofman94}. However, in the pressure model,
the growth rate of the perturbation depends on the scale. Therefore the
peculiar velocity distribution will deviate from Gaussian during
evolution. Of course, the peculiar velocity distribution in an N-body
simulation becomes non-Gaussian~\cite{ZQSW94}. Does the effect of
the pressure cause the occurrence of the non-Gaussian distribution? We think
that the time evolution of the peculiar velocity distribution is
one of the more interesting problems.

In our model, we introduce the strong simplification that the
velocity dispersion is approximately isotropic, i.e. the stress tensor
is diagonal and a pressure-like term~\cite{bdp}. However, in general,
the velocity dispersion does not remain isotropic in nonlinear
regime. Until when is the assumption to ignore anisotropic velocity
dispersion reasonable? Maartens et al.~\cite{mtm} discussed a relativistic
kinetic theory generalization which also incorporates
anisotropic velocity dispersion. Then they added these effects to the
linear development of density inhomogeneity and found exact solutions for
their evolution. In a Newtonian description, though the equations is not
generally closed, we will consider anisotropic velocity dispersion
and the higher-order velocity cumulant and estimate
their effects on the evolution of density inhomogeneity.

\begin{acknowledgments}
We would like to thank Thomas Buchert, Kei-ichi Maeda, Masaaki Morita, 
Momoko Suda, and Hideki Yahagi for useful discussion and comment in the work.
For usage of COSMICS and $P^3M$ codes,
we would like to thank Edmund Bertschinger and Alexander Shirokov.
We also thank the referee who gave proper comments
and advice for modifications.

This work was supported in part by a Waseda University Grant
for Special Research Projects (Individual Research 2002A-868 and
2003A-089).
\end{acknowledgments}


\end{document}